\documentclass[10pt, conference, letterpaper]{IEEEtran}
\IEEEoverridecommandlockouts

\usepackage{cite}
\usepackage{amsmath,amssymb,amsfonts}
\usepackage{graphicx}
\usepackage{booktabs}
\usepackage{subfig}
\usepackage{pifont}
\usepackage{url}

\def\BibTeX{{\rm B\kern-.05em{\sc i\kern-.025em b}\kern-.08em
    T\kern-.1667em\lower.7ex\hbox{E}\kern-.125emX}}

\begin{document}

\title{NetRL: Powering Network Decisions via NetGRPO-Driven Reinforcement Learning}

\author{\IEEEauthorblockN{Lingfeng Peng, Xiaohui Xie, Yang Bao, and Yong Cui}
\IEEEauthorblockA{Tsinghua University\\
\url{plf24@mails.tsinghua.edu.cn}, \url{xiexiaohui@tsinghua.edu.cn}\\
\url{baoy21@mails.tsinghua.edu.cn}, \url{cuiyong@tsinghua.edu.cn}}}

\maketitle

\begin{abstract}
Network decision-making tasks, including adaptive bitrate streaming, load balancing, and congestion control, require policies to process heterogeneous states and actions while balancing near- and long-term consequences. Existing solutions generalize poorly across network tasks, require substantial task-specific redesign, and lack reliable evaluation of long-horizon action consequences. Methods that learn through online interaction additionally face unstable optimization. We present \textbf{NetRL}, a two-stage LLM-based reinforcement learning framework that combines supervised fine-tuning on demonstration trajectories with online policy optimization. A universal mixed-embedding interface combines textual task descriptions with precise numerical states, while a network action head produces valid decisions. At its core, NetGRPO evaluates candidate actions over multiple future trajectories, directs deeper search toward uncertain alternatives through confidence-aware budgeting, and performs group-relative, clipped, and KL-regularized updates. NetRL achieves 23.6\% higher QoE than NetLLM, 12.2\% lower job completion time than Genet, and 24.1\% higher network utility than Orca, yielding state-of-the-art performance among the evaluated methods. Confidence-aware budgeting reduces end-to-end training time by 27.1\% without reducing validation QoE. NetRL also converges stably and generalizes to unseen network traces without retraining. Code and model weights are available at the repository linked in this paper.
\end{abstract}


\section{Introduction}
\label{sec:introduction}
Network decision-making---selecting actions from real-time observations to optimize network performance---encompasses critical tasks such as adaptive bitrate (ABR) streaming, load balancing (LB), and congestion control (CC) in modern communication infrastructures~\cite{10.1145/3098822.3098843,10113685,pmlr-v97-jay19a}. These tasks share a common structure: an agent observes a network state, selects a discrete action, and receives a scalar reward---yet each action can influence system states and performance over multiple subsequent decisions in non-stationary environments.

Existing approaches fall into three paradigms. Heuristic algorithms (e.g., BBA for ABR, Cubic for CC) use hand-crafted rules that are fast but brittle---every new condition demands expert re-engineering~\cite{10.1145/2619239.2626296,cubic2008}. Classic deep reinforcement learning (DRL) (e.g., Pensieve~\cite{10.1145/3098822.3098843}, Genet~\cite{10.1145/3544216.3544243}, Decima~\cite{10.1145/3341302.3342080}) learns from interaction but generalizes poorly across tasks, typically requiring entirely different model architectures for each new problem. Large language model (LLM)-based approaches~\cite{10.1145/3651890.3672268} achieve stronger performance through massive parameter counts, yet their task-specific interfaces underutilize pre-trained knowledge, their offline or frozen decision pipelines lack environment-grounded comparison of long-horizon action consequences, and direct online fine-tuning remains unstable. Taken together, these approaches leave three fundamental challenges unresolved:

\begin{enumerate}
    \item \textbf{Generalization Barrier (C1).} Network tasks differ fundamentally in what they observe, what they control, and how decisions cascade over time. ABR selects video quality from throughput and buffer states; LB dispatches jobs to minimize completion time; CC balances throughput against delay. These tasks embed fundamentally different control logic under heterogeneous state and action spaces.
    \item \textbf{Myopic Action Evaluation (C2).} Immediate feedback and a single sampled trajectory provide incomplete evidence of long-horizon consequences: an aggressive ABR bitrate may improve immediate quality but deplete the buffer and cause later rebuffering.
    \item \textbf{Instability Trap (C3).} Reinforcement learning is notorious for training instability~\cite{henderson2018deep}, while policy optimization for LLMs introduces additional stability challenges~\cite{zheng2023secretsrlhfppo}. Frozen-LLM approaches (Confucius~\cite{10.1145/3718958.3750537}, AgentRAN~\cite{agentran2025}) avoid training but forfeit the ability to improve.
\end{enumerate}

We propose \textbf{NetRL}\footnote{Code and model weights: \url{https://anonymous.4open.science/r/NetRLinfocom-1F1C/README.md}.}, a two-stage LLM-based RL framework that directly addresses these three challenges. NetRL performs Supervised Fine-Tuning (SFT) on demonstration trajectories for rapid domain adaptation, then refines the policy through online RL.

\begin{enumerate}
    \item \textbf{Universal Mixed-Embedding Interface (C1).} Text prompts describe each task's objective, leveraging the LLM's pre-trained knowledge, while numerical features precisely encode the network state. MLP encoders and the native tokenizer jointly project these heterogeneous inputs into a unified hidden space, and a network action head guarantees valid actions.
    \item \textbf{Confidence-Aware Trajectory Generation (C2).} NetGRPO replaces single-trajectory rollouts with multi-branch tree search. It evaluates all candidate root actions, directs deeper search toward alternatives whose confidence intervals still overlap, and aggregates retained path returns to produce more precise long-horizon Q-value estimates with less redundant computation.
    \item \textbf{NetGRPO: Stable Policy Updates (C3).} Building on the more precise action estimates from tree-structured generation, NetGRPO performs stable policy optimization via group-relative advantage normalization and KL-regularized updates, directly addressing the training instability of single-trajectory RL.
\end{enumerate}

We evaluate NetRL on ABR, LB, and CC against three categories of baselines: rule-based heuristics, classic DRL, and the LLM-based NetLLM. NetRL achieves 23.6\% higher QoE over NetLLM~\cite{10.1145/3651890.3672268} on ABR (0.89 vs.\ 0.72), 12.2\% lower JCT over Genet~\cite{10.1145/3544216.3544243} on LB (2.65 vs.\ 3.02), and 24.1\% higher utility over Orca~\cite{orca2020} on CC (0.36 vs.\ 0.29), yielding the best overall results among the evaluated baselines on all three tasks. These improvements are statistically significant across all baselines and tasks. NetRL further demonstrates strong generalization and stable convergence. On paired ABR runs, confidence-aware tree budgeting reduces end-to-end training time by 27.1\% without reducing validation QoE.

Our contributions are:
\begin{enumerate}
    \item We propose \textbf{NetRL}, a two-stage LLM-based RL framework that combines SFT for rapid domain adaptation with online RL to improve beyond the policies represented in the demonstration data across three diverse network decision tasks.
    \item We design a \textbf{universal mixed-embedding interface} that projects textual task descriptions and numerical network states into a unified hidden space, leveraging pre-trained LLM knowledge and MLP encoders to handle heterogeneous inputs, and employs a \textbf{network action head} to guarantee valid actions.
    \item We introduce \textbf{NetGRPO}, a tree-structured online RL algorithm that combines multi-path action evaluation with confidence-aware allocation of deeper search, then performs stable policy updates through group-relative advantage normalization and KL regularization.
    \item We conduct \textbf{extensive evaluation} across ABR, LB, and CC, demonstrating the best overall performance among the evaluated baselines, strong generalization, stable convergence, and reduced training cost through confidence-aware tree budgeting.
\end{enumerate}

\section{Background and Related Work}
\label{sec:background}
\subsection{Problem Formulation}

Network decision-making tasks are critical for the automation and intelligence of network operations~\cite{9069178}. They can generally be formulated as a Markov Decision Process (MDP) $\mathcal{M}=(\mathcal{S},\mathcal{A},\mathcal{P},r,\gamma)$. At each decision step $t$, an agent observes a state $s_t \in \mathcal{S}$, selects an action $a_t \in \mathcal{A}$ according to a parameterized policy $\pi_\theta(a_t|s_t)$, receives an immediate reward $r_t=r(s_t,a_t)$, and transitions to the next state according to $s_{t+1}\sim\mathcal{P}(\cdot|s_t,a_t)$. The objective is to learn policy parameters $\theta$ that maximize the expected cumulative discounted reward $J(\theta)=\mathbb{E}_{\pi_\theta,\mathcal{P}}[\sum_{t=0}^{\infty}\gamma^t r_t]$. NetRL focuses on network decision problems with finite action spaces, consistent with its enumeration of candidate actions during training.

\begin{figure}[t]
    \centering
    \includegraphics[width=0.70\linewidth]{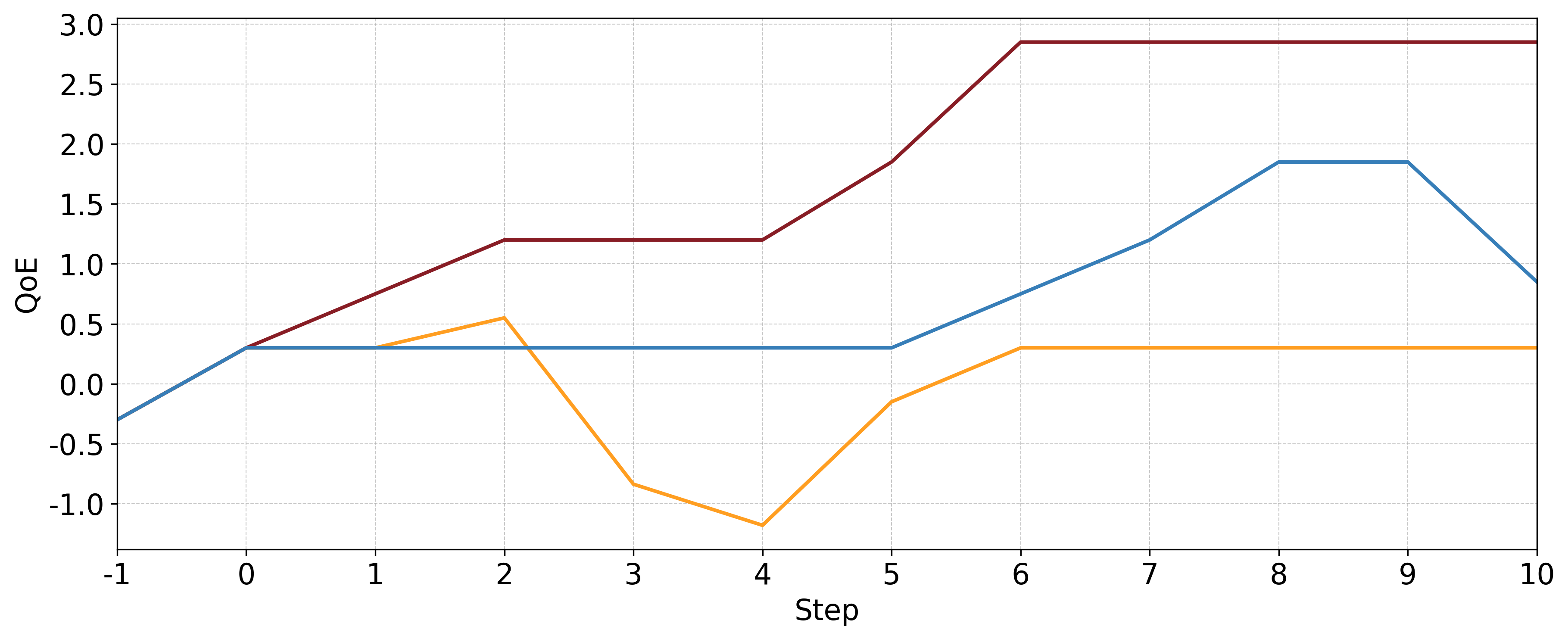}
    \caption{Long-horizon sensitivity in ABR. Starting from the same state, we take different bitrate actions at step~0 and then follow the same policy in all subsequent steps, resulting in substantially different cumulative QoE outcomes.}
    \label{fig:butterfly}
\end{figure}

ABR, LB, and CC instantiate this general formulation at different layers of the network stack. In general, \textit{ABR} selects the quality of successive video segments according to network and playback conditions, balancing visual quality, rebuffering, and quality variation~\cite{10.1145/3098822.3098843}. \textit{LB} assigns arriving workloads among available computing resources according to their service capacities and current loads, seeking low completion or response time and efficient resource utilization~\cite{10.5555/3454287.3454511}. \textit{CC} adapts sending behavior based on end-to-end network feedback to balance throughput, latency, packet loss, and fairness~\cite{pmlr-v97-jay19a}. Although these tasks share a sequential decision-making structure, their state representations, action semantics, and optimization objectives differ substantially. They also share long-horizon dependencies, as an action can affect network states and performance over multiple subsequent decision steps (Fig.~\ref{fig:butterfly}).

\subsection{Existing Approaches and Limitations}

\textbf{Heuristic Methods.} Heuristic methods use hand-crafted rules to map network observations to actions. Representative examples include BBA for ABR~\cite{10.1145/2619239.2626296}, Least-Work scheduling for LB, and Cubic for CC~\cite{cubic2008}. These methods are fast and interpretable, but their performance depends heavily on the expertise and foresight of the domain experts who design them. Once deployed, their fixed rules cannot learn from operational experience or autonomously adjust their decisions as network conditions evolve. Consequently, a heuristic designed for one task, environment, or objective often performs poorly when state distributions, constraints, or optimization objectives change. Adapting it to new conditions requires further expert re-engineering~\cite{10.1145/3603269.3604838}, resulting in limited generalizability.

\textbf{Classic Deep Reinforcement Learning.} DRL methods replace fixed rules with policies learned through environment interaction. Representative systems include Pensieve for ABR~\cite{10.1145/3098822.3098843}, Decima for cluster scheduling~\cite{10.1145/3341302.3342080}, Aurora for CC~\cite{pmlr-v97-jay19a}, and Genet for training policies under diverse network conditions~\cite{10.1145/3544216.3544243}. Although these methods can outperform heuristics within their target tasks, each task typically requires a separately designed state encoder, policy architecture, and training procedure, resulting in poor cross-task generalization. While DRL optimizes cumulative reward, conventional model-free methods estimate action values from a limited number of sampled trajectories. Delayed network effects therefore produce noisy and potentially unreliable long-horizon estimates. Their stochastic policy updates are also sensitive to random seeds and changing network conditions, resulting in unstable training and inconsistent final policies~\cite{henderson2018deep}.

\textbf{LLM-Based Methods.} LLMs have recently been applied to networking through three main paradigms. \textit{Offline code generation} methods such as NADA~\cite{10.1145/3696348.3696868} use an LLM to modify existing network algorithms. \textit{Offline imitation learning} methods such as NetLLM~\cite{10.1145/3651890.3672268} fine-tune an LLM on expert trajectories using task-specific multimodal encoders and action heads. \textit{Task orchestration and planning} methods such as Confucius~\cite{10.1145/3718958.3750537}, AgentRAN~\cite{agentran2025}, LMTE~\cite{lmte2026}, and Network CoPilot~\cite{networkcopilot2025} use frozen language models to generate workflows, plans, or configurations.

Despite their strong performance, these approaches have three limitations. First, their task interfaces generalize poorly. NetLLM requires task-specific encoders and training data, while code-generation and orchestration methods rely on domain-specific templates, tools, or vocabularies. Supporting a new task therefore requires substantial redesign and underutilizes the LLM's pre-trained knowledge.

Second, they do not explicitly compare the long-term consequences of alternative actions. Imitation learning supervises only the demonstrated action, while frozen planners do not use environment-grounded multi-path rollouts to evaluate candidate actions. Their decisions thus remain constrained by demonstrations, generated code, or predicted plans.

Third, they cannot stably improve through online interaction. Offline or frozen policies cannot learn from environment feedback, whereas directly optimizing a billion-parameter LLM with noisy, non-stationary network rewards requires careful policy constraints to maintain stable training~\cite{zheng2023secretsrlhfppo}.

Table~\ref{tab:comparison} summarizes these capability differences among representative LLM-based approaches.

\begin{table}[t]
    \centering
    \caption{Comparison of LLM-based network decision approaches.}
    \label{tab:comparison}
    \footnotesize
    \begin{tabular}{lccc}
        \toprule
        \textbf{Requirement} & \textbf{NetLLM} & \textbf{Confucius} & \textbf{NetRL} \\
        \midrule
        Common Interface Architecture & \ding{55} & \ding{55} & \ding{51} \\
        Multi-Path Action Evaluation   & \ding{55} & \ding{55} & \ding{51} \\
        Stable Online Policy Updates   & \ding{55} & \ding{55} & \ding{51} \\
        Closed-Loop Network Control    & \ding{51} & \ding{55} & \ding{51} \\
        \bottomrule
    \end{tabular}
\end{table}

\subsection{Remaining Challenges}

The limitations above reveal three requirements that remain unresolved by existing approaches.

\textbf{C1---Generalization Barrier.} Network decision tasks vary widely in their observation modalities, state representations, action spaces, control timescales, and optimization objectives. Existing methods typically address these differences with separately designed end-to-end pipelines, making representations and learned capabilities difficult to reuse across tasks. This heterogeneity limits cross-task generalization and motivates a common interface architecture that can support diverse network decision problems while preserving their task-specific semantics.

\textbf{C2---Myopic Action Evaluation.} Network actions can affect system states and performance over multiple subsequent decision steps, as illustrated in Fig.~\ref{fig:butterfly}. Immediate rewards provide no direct evidence of these delayed effects, while a single sampled trajectory captures only one possible future. Reliable action evaluation therefore requires comparing multiple possible downstream trajectories for each candidate action.

\textbf{C3---Instability Trap.} Unlike offline training on a fixed dataset, online RL continually changes the state--action distribution through policy-driven exploration and learns from noisy, delayed rewards, making its optimization substantially more prone to instability~\cite{PARK2025122081}. Policy optimization for LLMs introduces further stability challenges~\cite{zheng2023secretsrlhfppo}. Existing LLM-based methods therefore often avoid online learning by using frozen or offline policies, but consequently forfeit the ability to improve through environment interaction. A stable mechanism for online LLM policy optimization is thus required.

\section{NetRL Design}
\label{sec:design}
\begin{figure*}[t]
    \centering
    \includegraphics[width=0.8\textwidth]{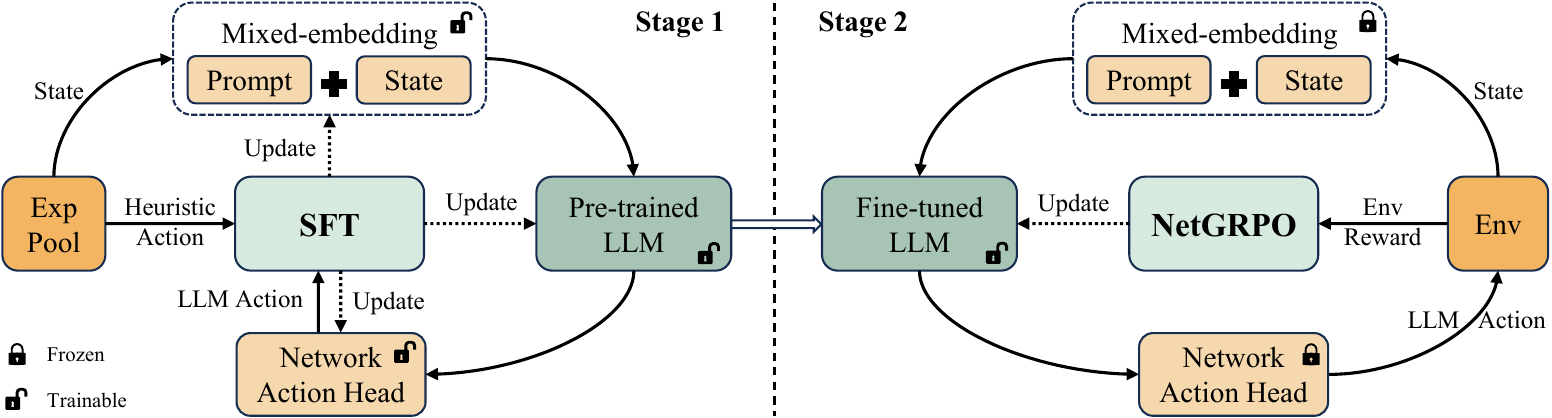}
    \caption{An overview of the NetRL framework.}
    \label{fig:workflow}
\end{figure*}

We present \textbf{NetRL}, a two-stage framework for adapting pre-trained LLMs to heterogeneous network decision tasks. As illustrated in Fig.~\ref{fig:workflow}, NetRL first performs Supervised Fine-Tuning (SFT) on demonstration trajectories to establish a competent initial policy, and then improves the policy through online interaction with the network environment. Its design directly addresses the three challenges identified in Section~\ref{sec:background}. The universal mixed-embedding interface provides a common architecture for processing heterogeneous network observations and producing valid task-specific actions (C1). NetGRPO combines multi-path evaluation with confidence-aware allocation of deeper tree search to estimate long-horizon action consequences efficiently (C2), then uses group-relative advantage normalization and KL-regularized updates to stabilize online policy optimization (C3).

\subsection{Two-Stage Training Paradigm}

Network decision tasks present a tradeoff between efficient policy initialization and experience-driven improvement. Starting online RL from a pre-trained LLM without a task-adapted policy is sample-inefficient and unstable, because the model must simultaneously learn the task-specific state--action mapping and optimize long-term control performance. Conversely, offline imitation provides an effective initialization but remains bounded by the quality and coverage of its demonstrations. NetRL combines the two approaches to obtain both rapid adaptation and further performance improvement.

\textbf{Stage 1: Supervised Fine-Tuning.} We construct an offline dataset of demonstration trajectories by running established policies in each target environment. The pre-trained LLM undergoes SFT on these trajectories using low-rank adaptation (LoRA)~\cite{hu2021lora}, which preserves pre-trained knowledge while efficiently adapting to the target domain. This stage teaches the model the task-specific state--action mapping and produces a competent initial policy without requiring costly online exploration.

\textbf{Stage 2: Online Reinforcement Learning.} The SFT-initialized policy is then further optimized through direct interaction with the training environment. By observing environment states, taking actions, and receiving rewards, the policy explores beyond the demonstration data and directly optimizes cumulative return. Unlike imitation learning, which only reproduces demonstrated actions, this interaction allows the policy to improve based on the consequences of its own decisions. However, delayed action effects and stochastic network dynamics make rewards from individual trajectories inherently noisy. NetGRPO addresses this problem through multi-path action evaluation and group-relative policy updates (Section~\ref{subsec:netgrpo}). Here, \emph{online} refers to environment interaction during policy training; deployment uses a fixed policy and requires only a single forward pass.

\subsection{Universal Mixed-Embedding Interface}
\label{subsec:interface}

\textbf{Motivation.} Network decision tasks differ substantially in their objectives, observation semantics, state dimensions, and action spaces. Existing approaches typically accommodate these differences through separately designed input encoders and output modules, making the resulting policy architecture difficult to reuse across tasks. Moreover, directly converting numerical measurements into text may lose precision, whereas using only numerical features prevents the model from explicitly understanding the task objective and feature semantics. NetRL therefore requires a common interface that combines the semantic expressiveness of natural language with the precision of numerical representations.

\begin{figure}[t]
    \centering
    \includegraphics[width=0.80\linewidth]{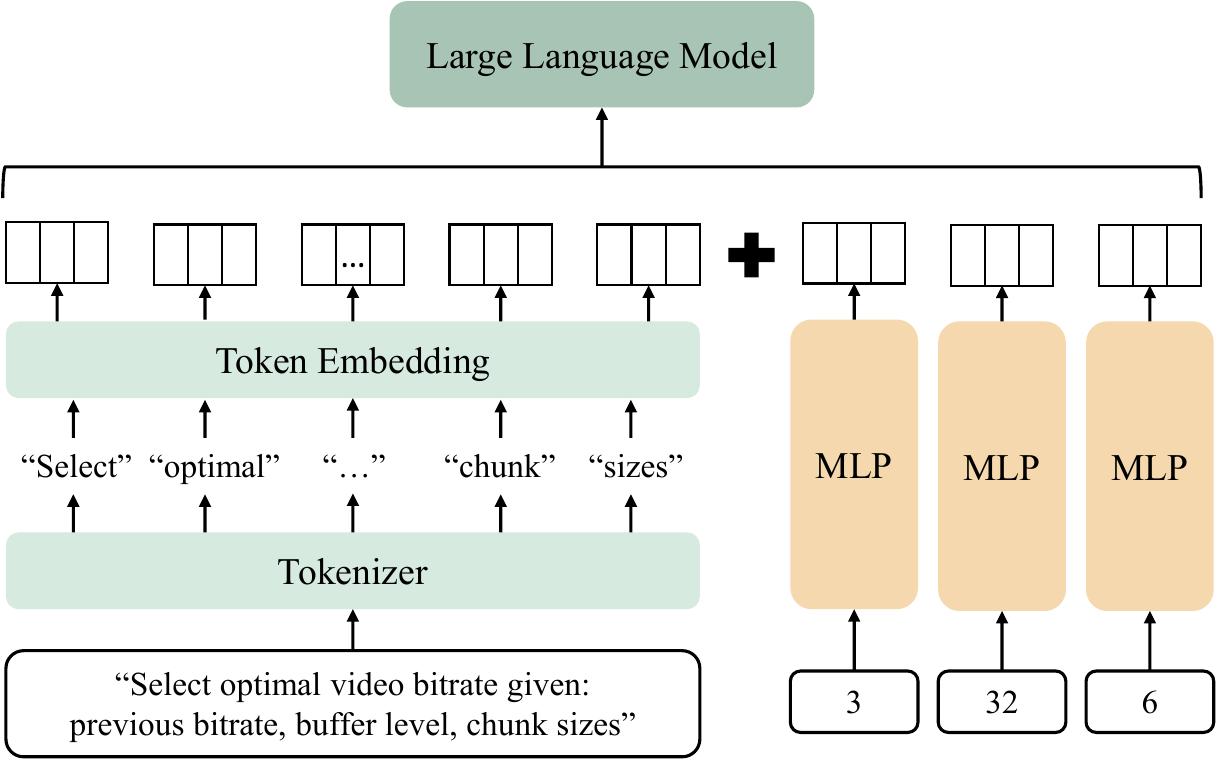}
    \caption{The mixed-embedding interface. A textual prompt describing the task objective and feature semantics is processed by the native tokenizer, while numerical state features are projected into the LLM hidden space through MLP encoders. The text and numerical embeddings are concatenated and jointly processed by the LLM.}
    \label{fig:mix-embed}
\end{figure}

\textbf{Mixed-Embedding Input.} At each decision step, the interface receives two complementary inputs: a textual prompt and a numerical network state. The prompt describes the task objective and the semantics of the observed features, and is processed by the LLM's native tokenizer and token-embedding layer. Each numerical feature is mapped directly into the LLM hidden space through an MLP-based encoder, preserving its continuous value without stringification or discretization. The resulting text and numerical embeddings are concatenated into a single input sequence and processed jointly by the LLM, as illustrated in Fig.~\ref{fig:mix-embed}. This design allows the policy to use pre-trained language knowledge to interpret the task description while retaining precise information about the current network state.

\textbf{Network Action Head.} The standard language-modeling head is replaced by a network action head that maps the final hidden representation to logits over the valid action set. Applying a softmax produces the categorical policy $\pi_\theta(a_t|s_t)$ used by SFT and online RL. The action head prevents invalid textual outputs and ensures that every decision corresponds to an admissible network action. Its output dimension is configured according to the action space of the target task.

\textbf{Why This Addresses the Generalization Barrier.} NetRL reuses the same prompt--embedding--backbone--action-head architecture across tasks; adapting it to a new task requires specifying the task prompt, numerical features, and valid action set rather than redesigning the complete policy architecture. Text prompts encode task objectives and feature semantics, MLP embeddings preserve numerical precision, and the action head enforces task-valid outputs. This design is supported by the ablation results in Section~\ref{sec:evaluation}: replacing numerical embeddings with stringified inputs degrades performance across all three tasks.

\subsection{NetGRPO Algorithm}
\label{subsec:netgrpo}

\textbf{Motivation.} After SFT, NetRL improves its policy through environment interaction. However, a single sampled trajectory reveals only one possible consequence of an action, making long-horizon action values sensitive to stochastic network dynamics (C2). Directly updating a large policy from these noisy estimates can further produce unstable optimization (C3). NetGRPO addresses the two problems jointly: tree-structured trajectory generation evaluates each candidate action over multiple future paths, confidence-aware budgeting avoids spending the same search depth on already clear and still ambiguous choices, and group-relative policy optimization converts the resulting action values into normalized and regularized policy updates.

\begin{figure*}[t]
    \centering
    \includegraphics[width=0.9\textwidth]{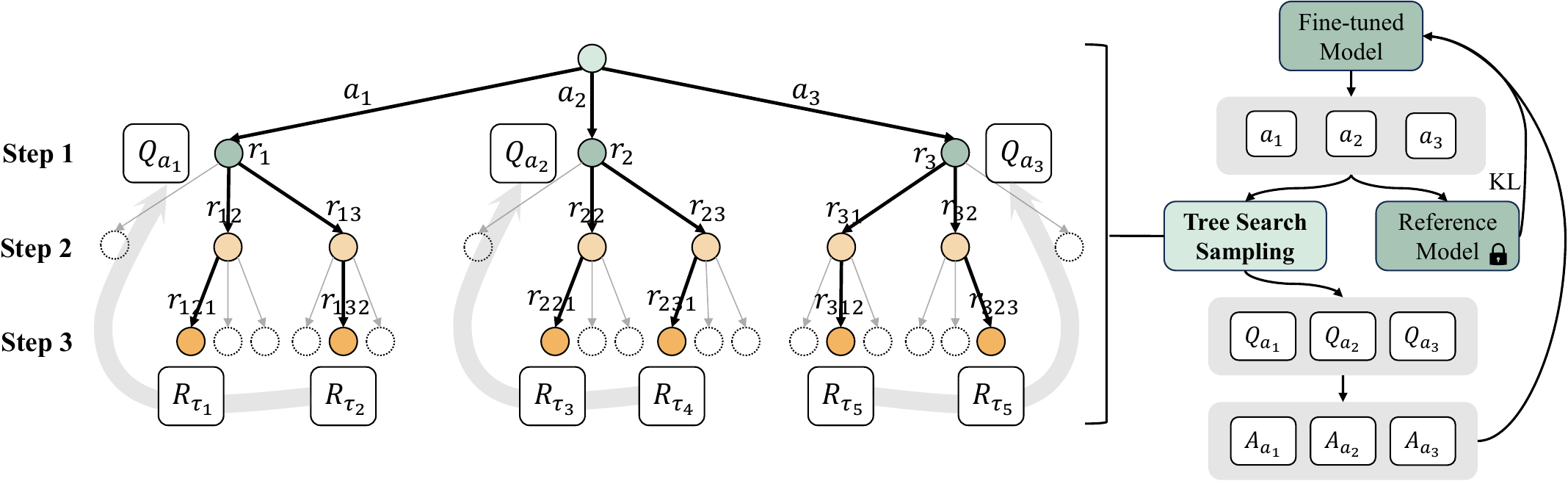}
    \caption{The NetGRPO algorithm. All candidate actions are expanded from the root state. Descendant trajectories are generated through pruned beam rollout, and their cumulative returns are aggregated to estimate the long-term value of each root action. The resulting action group is used for group-relative policy optimization.}
    \label{fig:netGRPO}
\end{figure*}

\textbf{Algorithm overview.} At each policy update, NetGRPO starts from a root state and evaluates all admissible root actions under the same network condition. Each root action initializes one branch, after which the current policy interacts with independently evolved environment states to generate multiple possible futures. A shared shallow search provides evidence for every action, and confidence-aware budgeting either stops the tree when the choice is clear or assigns deeper rollouts only to unresolved alternatives. The environment provides the original task reward at every transition; NetGRPO introduces no auxiliary reward function. It aggregates retained path returns into an action-value estimate for every root action and optimizes the policy using their relative values within the complete root-state action group.

\subsubsection{\textbf{Tree-Structured Multi-Path Evaluation}}

Given a root state $s_t$, NetGRPO constructs a finite-depth trajectory tree under the current policy $\pi_\theta$. Unlike a single rollout, the tree evaluates multiple possible futures for every candidate action.

\textbf{Full root expansion.} NetGRPO first enumerates every action $a_i\in\mathcal{A}$ at the root. Executing $a_i$ produces the next state $s_{t+1}^{(i)}$ and immediate reward $r(s_t,a_i)$. Each root action is retained and forms one member of the action group used for the subsequent policy update.

\textbf{Pruned beam rollout.} For depths $d>1$, the states of all current leaf nodes are batched and passed through $\pi_\theta$. For each parent node $v$, all candidate successor actions are evaluated. A child node $v'$ generated by action $a$ stores the cumulative discounted return
\begin{equation}
    R(v') = R(v) + \gamma^{d-1}r(s_v,a),
    \label{eq:path_return}
\end{equation}
where $R(v_{\mathrm{root}})=0$ and $s_v$ is the state represented by node $v$.

Expanding every successor causes the number of trajectories to grow exponentially with depth. NetGRPO therefore retains the top-$k_d$ children of each parent according to
\begin{equation}
    \mathrm{Score}(v,a) = \bar{R}(v') + C_d\pi_\theta(a|s_v),
    \label{eq:beam_score}
\end{equation}
where $\pi_\theta(a|s_v)$ is the current policy probability. Among the sibling candidates of the same parent, we normalize the cumulative return as $\bar{R}(v')=(R(v')-R_{\min})/(R_{\max}-R_{\min})$, where $R_{\min}$ and $R_{\max}$ are their minimum and maximum returns; when all sibling returns are equal, we set $\bar{R}(v')=0.5$. Both the beam width $k_d$ and policy coefficient $C_d$ decrease with depth. We use $k_d=[4,3,2]$ and $C_d=[3.0,2.0,1.5]$ for depths $d=[2,3,4]$. This schedule preserves broader policy-guided exploration near the root and increasingly favors high-return paths at deeper levels. All root actions remain in the tree; pruning and budget allocation affect only their descendants.

\textbf{Confidence-aware tree budgeting.} After depth 3, a logistic calibrator uses the value margin and policy uncertainty to estimate whether deeper search will change the best root action. High-confidence states stop further tree expansion. Otherwise, a validation-fitted ridge model predicts each root action's full-depth value $\widetilde{Q}_i$ and uncertainty $u_i$ from its depth-3 frontier. NetGRPO extends only actions satisfying $\widetilde{Q}_i+\lambda u_i \geq \max_j(\widetilde{Q}_j-\lambda u_j)-m$, where $\lambda$ and margin $m$ are selected on disjoint validation traces. Non-extended actions retain calibrated estimates capped below the best expanded value, so every action remains in the group while deeper computation focuses on unresolved alternatives.

\textbf{Path-aggregated Q-value estimation.} Let $\mathcal{T}(a_i)$ denote the retained leaf paths whose first action is $a_i$. NetGRPO estimates the value of each root action by averaging the cumulative returns of its paths:
\begin{equation}
    \hat{Q}(s_t,a_i) = \frac{1}{|\mathcal{T}(a_i)|}\sum_{\tau\in\mathcal{T}(a_i)}R(\tau).
    \label{eq:q_estimate}
\end{equation}
Here, $R(\tau)$ is the discounted sum of the original environment rewards along path $\tau$, rather than a separately designed ``NetGRPO reward.'' When the complete tree stops at depth 3, $\mathcal{T}(a_i)$ contains the retained depth-3 paths. Under selective depth-4 expansion, expanded actions use their depth-4 path returns, while non-extended actions use the calibrated $\widetilde{Q}_i$ described above. Consequently, $\hat{Q}(s_t,a_i)$ summarizes the multi-step outcomes of taking $a_i$ across the retained policy-guided paths. Compared with assigning an action the return of only one sampled rollout, aggregating multiple retained paths reduces dependence on a particular stochastic future. Full root expansion also ensures that all candidate actions are compared under the same initial network state, instead of leaving some actions unevaluated because they were not sampled. Because beam pruning may exclude delayed-benefit paths, NetGRPO does not assume that this estimate is unbiased.

\subsubsection{\textbf{Group-Relative Policy Optimization}}

The action-value pairs $\{(a_i,\hat{Q}(s_t,a_i))\}_{i=1}^{|\mathcal{A}|}$ form an action group $G_t$. NetGRPO normalizes their values within the group:
\begin{equation}
    \hat{A}(a_i) =
    \frac{\hat{Q}(s_t,a_i)-\mu_Q}{\sigma_Q},
    \quad \sigma_Q \ge 10^{-5},
    \label{eq:group_advantage}
\end{equation}
where $\mu_Q$ and $\sigma_Q$ are the group mean and standard deviation; when $\sigma_Q<10^{-5}$, we set all advantages in the group to zero. Actions expanded from the same root state form a natural comparison group because they share the same initial network condition. Their normalized advantages therefore indicate which actions are relatively preferable at that state, without requiring an additional learned value function as in actor--critic PPO. This is particularly useful for heterogeneous network environments, where absolute reward magnitudes may vary substantially across states and operating conditions. NetGRPO still retains the PPO-style clipped objective to bound each policy update.

Let $\rho_i(\theta)=\pi_\theta(a_i|s_t)/\pi_{\theta_{\mathrm{old}}}(a_i|s_t)$. The clipped surrogate for action $a_i$ is
\begin{equation}
\begin{aligned}
    L_i^{\mathrm{CLIP}}(\theta)
    = \min\big(&\rho_i(\theta)\hat{A}(a_i),\\
      &\mathrm{clip}(\rho_i(\theta),1-\epsilon,1+\epsilon)
       \hat{A}(a_i)\big).
\end{aligned}
\end{equation}
NetGRPO updates the policy by maximizing
\begin{equation}
\begin{aligned}
    \mathcal{J}_{\mathrm{NetGRPO}}(\theta)
    = \mathbb{E}_{s_t\sim\mathcal{D}}\bigg[
      &\frac{1}{|\mathcal{A}|}\sum_{a_i\in G_t}L_i^{\mathrm{CLIP}}(\theta)\\
      &-\beta D_{\mathrm{KL}}(\pi_{\mathrm{ref}}\|\pi_\theta)\bigg],
\end{aligned}
    \label{eq:netgrpo_objective}
\end{equation}
where $\mathcal{D}$ denotes the distribution of root states collected through online interaction, $\pi_{\theta_{\mathrm{old}}}$ is the policy before the current update, and $\pi_{\mathrm{ref}}$ is the frozen SFT policy. Group normalization provides a state-relative learning signal, clipping limits abrupt policy changes, and the KL penalty anchors online updates to the SFT initialization. Together, these mechanisms improve optimization stability without discarding the knowledge acquired during SFT.

\textbf{Mitigating Myopic Action Evaluation and Unstable Online Optimization.} The components of NetGRPO play distinct but complementary roles. Tree-structured rollout addresses myopic evaluation by tracing each root action over multiple steps and retained futures; confidence-aware budgeting preserves the complete action comparison while concentrating deeper computation where the available evidence cannot yet separate alternatives. Group-relative normalization converts these action values into state-relative learning signals, while clipping and KL regularization prevent noisy estimates from inducing abrupt changes to the large policy. Thus, multi-path evaluation improves the optimization target, adaptive budgeting controls its cost, and normalized, regularized updates improve how it is incorporated into the policy. Section~\ref{sec:evaluation} evaluates all three effects through final performance, paired cost measurements, and training dynamics.

\subsubsection{\textbf{Training Cost and Inference}}

Tree generation is used only during training. At inference, NetRL selects an action with a single policy forward pass and performs no tree rollout. For a six-action problem with depth $D=4$, exhaustive expansion evaluates $1{,}554$ non-root transitions. With $k_d=[4,3,2]$, NetGRPO evaluates at most 618 candidate transitions and produces 144 leaf paths; confidence-aware budgeting reduces a clear state to 186 transitions and selectively extends uncertain root branches otherwise.

This increases the training time per iteration from approximately 16\,ms for vanilla GRPO to 870\,ms, but does not add inference latency. Despite the higher per-iteration cost, more stable policy updates reduce oscillations and allow NetRL to converge to strong performance within 2--3 hours of online training, as shown in Section~\ref{sec:evaluation}.

\section{Evaluation}
\label{sec:evaluation}
We evaluate NetRL on three representative network decision tasks---ABR, LB, and CC. Our experiments answer four questions: (1) how NetRL compares with rule-based, classic DRL, and LLM-based methods; (2) whether it generalizes to unseen network settings; (3) how its main components affect performance and training stability; and (4) what computational cost it introduces. Unless otherwise stated, results report the mean and standard deviation over five random seeds. Error bars show $\pm 1$ standard deviation, while *, **, and *** denote $p{<}0.05$, $p{<}0.01$, and $p{<}0.001$, respectively, under Welch's t-test comparing NetRL with each baseline. Each evaluation seed instantiates a different network environment, so the resulting variation reflects robustness across diverse network conditions rather than differences among independent training runs.

\subsection{Experimental Setup}

\textbf{Tasks and Environments.} All experiments are simulation-based and follow established network benchmarks. ABR uses the Pensieve simulator~\cite{10.1145/3098822.3098843}, the Envivio-Dash3 videos, and FCC broadband traces~\cite{fcc2016}, with 235, 150, and 100 disjoint traces for training, validation, and testing, respectively. LB uses the Park simulator~\cite{10.5555/3454287.3454511} with Poisson job arrivals and Pareto-distributed job sizes. CC uses the Genet environment~\cite{10.1145/3544216.3544243} under varying bandwidth, delay, and loss conditions.

\textbf{Metrics.} For each task, we report one overall-performance metric and one robustness metric: average QoE and bad quality ratio for ABR, average Job Completion Time (JCT) and 99th-percentile waiting time for LB, and network utility and 99th-percentile round-trip time (RTT) for CC. The bad quality ratio is the fraction of decisions with a QoE score below zero, while network utility follows Genet's definition~\cite{10.1145/3544216.3544243}, balancing throughput against excess delay.

\textbf{Baselines.} We compare NetRL with rule-based heuristics, including BBA~\cite{10.1145/2619239.2626296}, MPC~\cite{10.1145/2785956.2787486}, BOLA~\cite{spiteri2016bola}, Round Robin, Least Work, Cubic~\cite{cubic2008}, and BBR~\cite{cardwell2016bbr}; classic DRL methods, including PPO~\cite{schulman2017ppo}, Genet~\cite{10.1145/3544216.3544243}, Decima~\cite{10.1145/3341302.3342080}, and Orca~\cite{orca2020}; and the LLM-based NetLLM~\cite{10.1145/3651890.3672268}. For a controlled comparison, NetLLM and NetRL use the same Qwen3-4B backbone.

\textbf{Implementation.} NetRL uses Qwen3-4B~\cite{yang2025qwen3technicalreport} with the interface described in Section~\ref{subsec:interface}. We train a separate policy for each task while reusing the same foundation model and interface architecture, with task-configured numerical embeddings and action heads. SFT demonstrations are collected from MPC, Decima, and BBR for ABR, LB, and CC, respectively. We use LoRA~\cite{hu2021lora} ranks 8 and 4 for SFT and NetGRPO. NetGRPO uses depth $D{=}4$, pruning widths $k_d{=}[4,3,2]$, policy coefficients $C_d{=}[3.0,2.0,1.5]$, $\gamma{=}0.99$, $\epsilon{=}0.1$, and $\beta{=}0.2$. For ABR, confidence calibrators and thresholds are selected only on validation traces. Training takes approximately 1 hour for SFT and 2--3 hours for NetGRPO on four NVIDIA A100 GPUs.

\subsection{Overall Performance}

We evaluate NetRL against all baselines on ABR, LB, and CC. Across all three tasks, NetRL achieves the best overall performance among the evaluated methods and consistently strong results on the complementary robustness metrics.

\begin{figure}[t]
    \centering
    \subfloat[Average QoE]{\label{fig:abr_performance_a}\includegraphics[width=0.48\columnwidth]{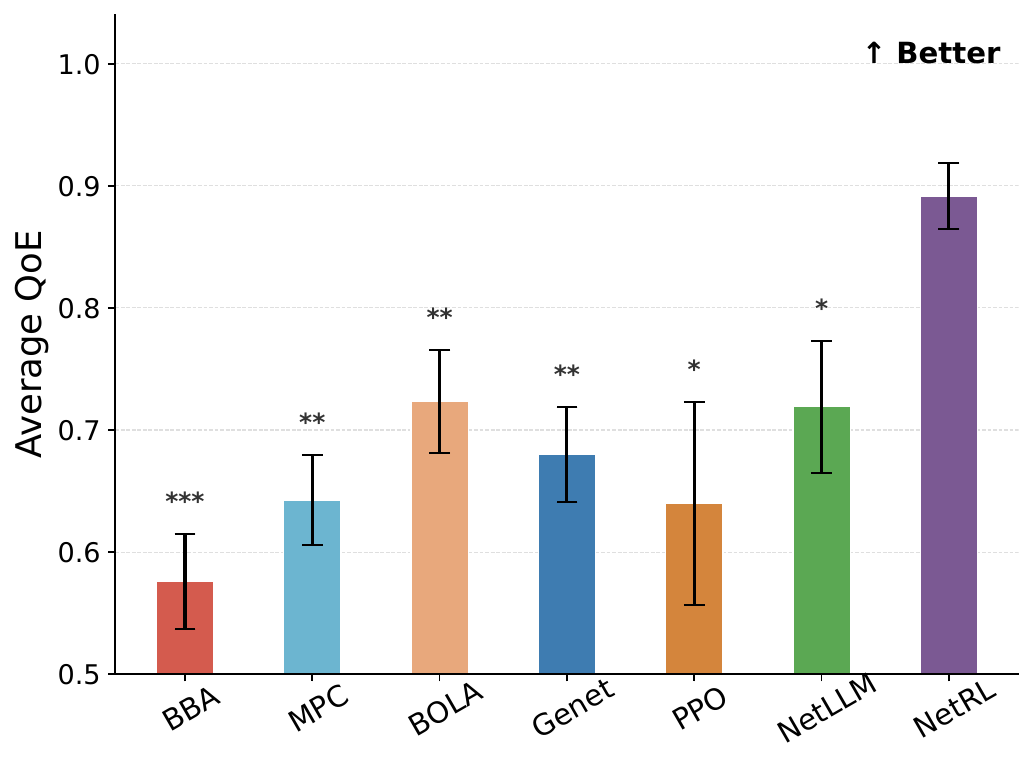}}%
    \hfill
    \subfloat[Bad Quality Ratio]{\label{fig:abr_performance_b}\includegraphics[width=0.48\columnwidth]{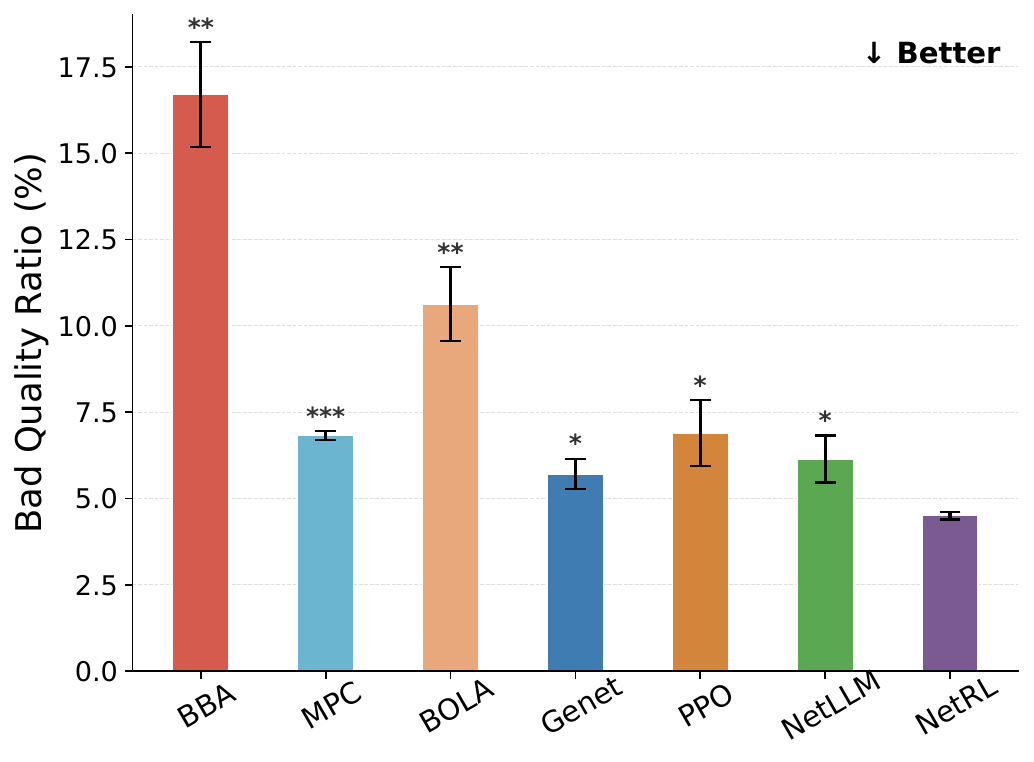}}%
    \caption{ABR performance.}
    \label{fig:abr_performance}
\end{figure}

\textit{ABR results.} Fig.~\ref{fig:abr_performance} presents the ABR results in terms of average QoE and bad quality ratio. BOLA reaches a competitive average QoE of 0.72, but its fixed rules adapt poorly when network conditions deteriorate. Genet and PPO learn through interaction, yet limited sampled trajectories can produce short-sighted estimates of delayed buffer and rebuffering effects. NetLLM also reaches an average QoE of 0.72 through its large model capacity and imitation learning, but may remain constrained by the coverage of its offline demonstrations.

NetRL achieves the best result on both metrics, with an average QoE of 0.89 and a bad quality ratio of 4.50\%. This improves QoE by 23.6\% over NetLLM and 23.3\% over BOLA, while reducing the bad quality ratio by 26.8\% relative to NetLLM and 73.1\% relative to BBA. Since NetLLM and NetRL use the same Qwen3-4B backbone, the improvement cannot be attributed to model scale. Instead, NetGRPO provides more precise action evaluation by evaluating each bitrate action over multiple future playback trajectories. This allows NetRL to account for delayed buffer and rebuffering effects rather than relying on a single sampled future.

\begin{figure}[t]
    \centering
    \subfloat[Average JCT]{\label{fig:lb_performance_a}\includegraphics[width=0.48\columnwidth]{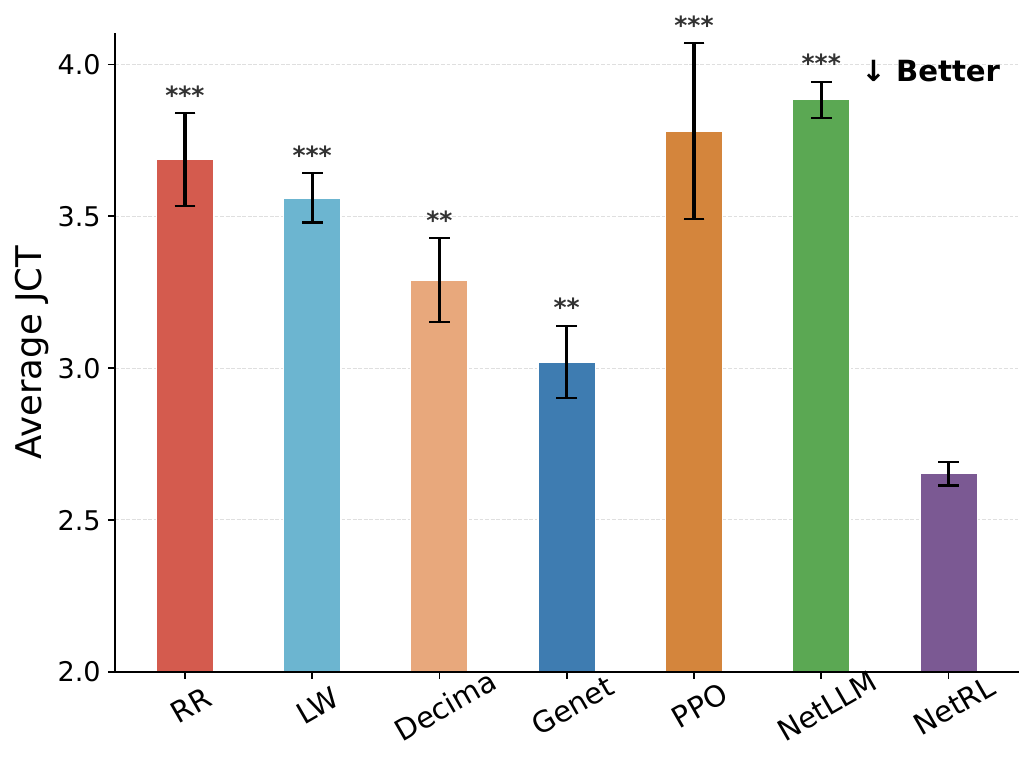}}%
    \hfill
    \subfloat[p99 Waiting Time]{\label{fig:lb_performance_b}\includegraphics[width=0.48\columnwidth]{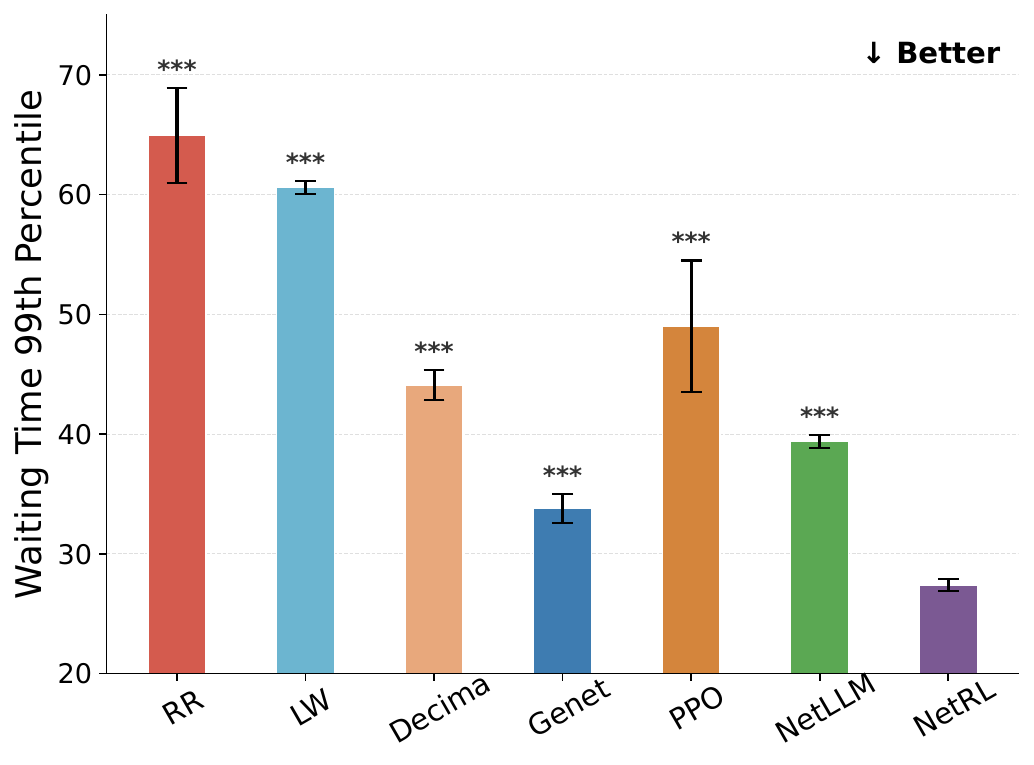}}%
    \caption{LB performance.}
    \label{fig:lb_performance}
\end{figure}

\textit{LB results.} Fig.~\ref{fig:lb_performance} presents the LB results in terms of average JCT and 99th-percentile waiting time. Round Robin and Least Work provide efficient rule-based baselines but cannot adapt their fixed decisions to changing workloads. Genet is the strongest learned baseline, achieving an average JCT of 3.02 and a 99th-percentile waiting time of 33.79. PPO and NetLLM perform particularly poorly, suggesting that their current input designs are not well suited to the LB state structure. PPO relies on a conventional numerical encoder, while NetLLM uses a manually designed task-specific encoder; neither effectively represents the coupled job and server states required for dispatch decisions.

NetRL reduces the average JCT to 2.65, outperforming Genet by 12.2\% and PPO by 29.9\%. It also lowers the 99th-percentile waiting time to 27.39, improving over Genet by 18.9\% and PPO by 44.1\%. The mixed-embedding interface preserves precise numerical workload states while using the textual prompt to encode the task objective and feature semantics. It therefore provides a more informative state representation without redesigning a task-specific encoder, allowing NetRL to maintain strong average and tail performance on LB.

\begin{figure}[t]
    \centering
    \subfloat[Network Utility]{\label{fig:cc_performance_a}\includegraphics[width=0.48\columnwidth]{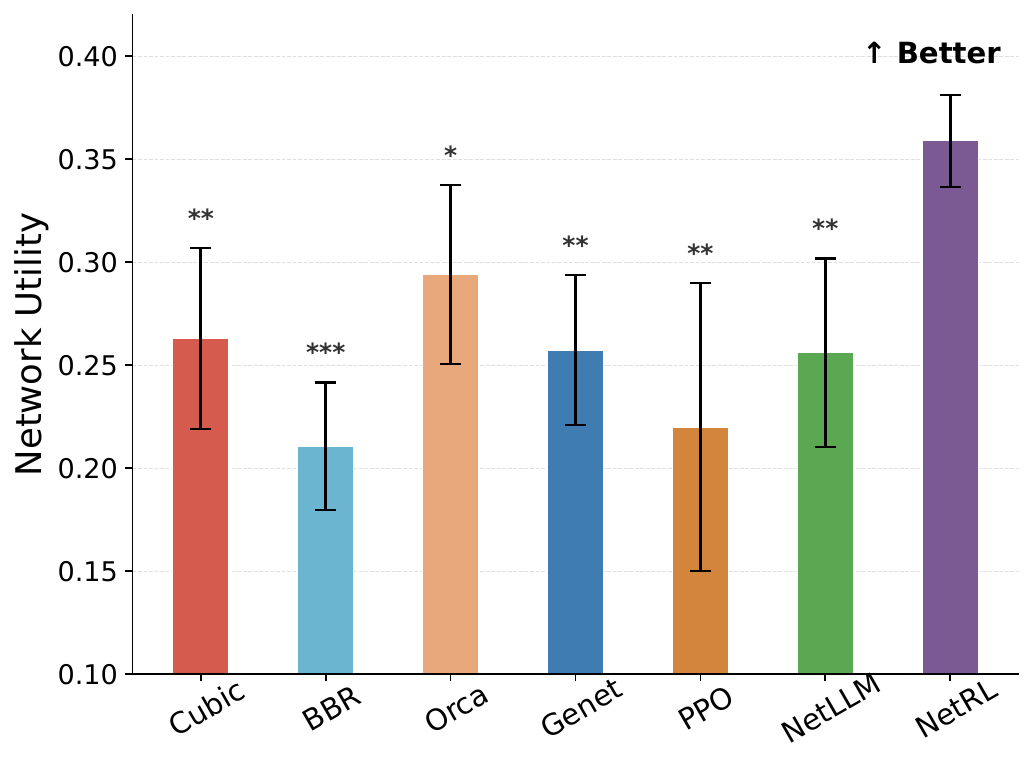}}%
    \hfill
    \subfloat[Tail Latency]{\label{fig:cc_performance_b}\includegraphics[width=0.48\columnwidth]{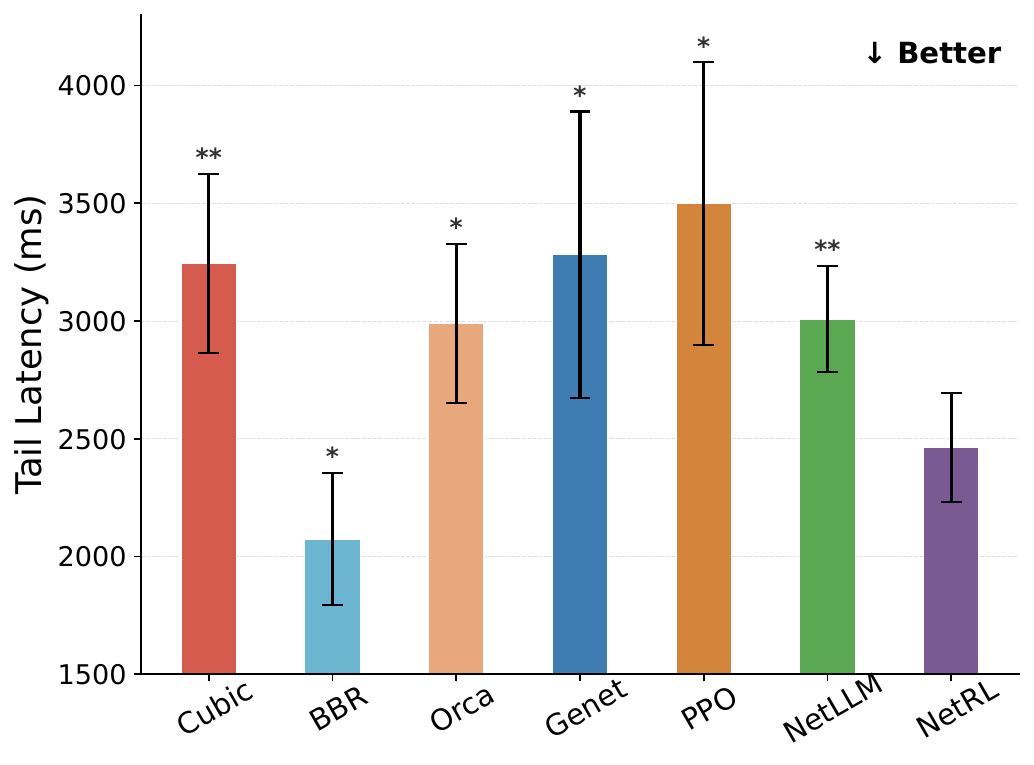}}%
    \caption{CC performance.}
    \label{fig:cc_performance}
\end{figure}

\textit{CC results.} Fig.~\ref{fig:cc_performance} presents the CC results in terms of network utility and 99th-percentile RTT. Cubic and BBR use fixed rules designed for different operating objectives. BBR achieves the lowest tail RTT of 2,074\,ms but a substantially lower network utility of 0.21, reflecting its conservative preference for limiting queue buildup over fully utilizing available bandwidth.

Learned CC methods face unstable optimization under rapidly changing bandwidth, delay, and loss. Delayed network feedback produces high-variance action estimates for conventional RL, while NetLLM's offline imitation learning is constrained by the coverage of its demonstrations and the difficulty of adapting a large-parameter model. Because NetLLM does not interact with the environment during training, it cannot use observed network outcomes to improve beyond the demonstrated behavior. These factors may limit final performance and lead to greater variation across diverse network environments.

NetRL achieves the highest network utility of 0.36, outperforming Orca by 24.1\% and both Genet and NetLLM by 38.5\%. Its 99th-percentile RTT is 2,462\,ms, which is 18.2\% lower than NetLLM and 24.1\% lower than Cubic. Although BBR has a lower tail RTT, it obtains this result by sacrificing substantial usable bandwidth. NetGRPO aggregates multiple future paths to reduce action-value variance, while group-relative normalization and KL regularization constrain abrupt policy changes. The small error bars further show that NetRL maintains consistent performance across diverse network environments, demonstrating strong robustness while achieving a better throughput--latency balance than the learned baselines.

Overall, the three tasks expose complementary advantages of NetRL: multi-path evaluation improves delayed action estimation in ABR (C2), the mixed-embedding interface provides effective reusable representations in LB (C1), and regularized online updates stabilize policy optimization in CC (C3). Together, these mechanisms enable strong overall and robustness performance across heterogeneous network decision tasks.

\subsection{Ablation Study}

We evaluate six controlled variants to quantify the contributions of NetRL's pre-trained backbone, universal interface, two-stage training paradigm, and NetGRPO algorithm. Table~\ref{tab:ablation} reports their performance across all three tasks.

\begin{table}[t]
    \centering
    \caption{Ablation results (ABR: QoE $\uparrow$; LB: JCT $\downarrow$; CC: utility $\uparrow$).}
    \label{tab:ablation}
    \footnotesize
    \begin{tabular}{lccc}
        \toprule
        \textbf{Variant} & \textbf{ABR} & \textbf{LB} & \textbf{CC} \\
        \midrule
        NetRL (full)                 & \textbf{0.89} & \textbf{2.65} & \textbf{0.36} \\
        \midrule
        w/o LLM Pre-training         & 0.67 & 3.57 & 0.28 \\
        w/o Mixed-Embedding          & 0.70 & 4.83 & 0.27 \\
        w/o Network Action Head      & 0.76 & 4.56 & 0.32 \\
        \midrule
        w/o SFT                      & 0.65 & 2.91 & 0.29 \\
        w/o RL (SFT only)            & 0.82 & 3.83 & 0.33 \\
        \midrule
        w/o NetGRPO (vanilla GRPO)   & 0.84 & 2.71 & 0.34 \\
        \bottomrule
    \end{tabular}
\end{table}

\textit{Pre-trained knowledge and universal interface.} In the \textit{w/o LLM Pre-training} variant, the backbone is randomly initialized instead of loading pre-trained weights. This reduces ABR QoE from 0.89 to 0.67, increases LB JCT from 2.65 to 3.57, and decreases CC utility from 0.36 to 0.28. The results confirm that pre-trained initialization provides useful representations for network decision-making. The textual prompt in NetRL's universal interface uses the pre-trained knowledge to describe the task objective and feature semantics.

The interface components are also necessary for translating this knowledge into effective network decisions. Replacing mixed embeddings with stringified numerical inputs reduces ABR QoE to 0.70 and increases LB JCT to 4.83 because precise numerical state information is lost. Without the network action head, the original language-modeling head generates action tokens that are mapped to network actions; this variant also reduces performance, especially on LB, where JCT increases to 4.56. Together, the textual prompt, numerical embeddings, and network action head allow the same interface architecture to interpret heterogeneous task semantics, preserve numerical precision, and produce valid actions.

\textit{Two-stage training.} Removing SFT reduces performance to 0.65 QoE on ABR, 2.91 JCT on LB, and 0.29 utility on CC, showing that demonstration-guided initialization is important for effective online learning. Using SFT alone (\textit{w/o RL}) also remains below the full model, particularly on LB, where JCT increases from 2.65 to 3.83. Thus, SFT provides a competent initial policy, while online interaction allows NetRL to improve beyond the demonstrated behavior. Although BBR supplies the CC demonstrations, SFT alone reaches only 0.33 utility, whereas online optimization further improves it to 0.36.

\textit{NetGRPO.} Replacing NetGRPO with vanilla GRPO while retaining the same pre-trained backbone reduces performance on every task: ABR QoE decreases from 0.89 to 0.84, LB JCT increases from 2.65 to 2.71, and CC utility decreases from 0.36 to 0.34. Vanilla GRPO uses the same computational resources and is trained for a comparable amount of time. This controlled comparison shows that NetGRPO consistently improves performance over vanilla GRPO under the same pre-trained backbone.

\subsection{Training Stability}

We track ABR QoE across training checkpoints to examine online-training stability.

\begin{figure}[t]
    \centering
    \includegraphics[width=0.56\linewidth]{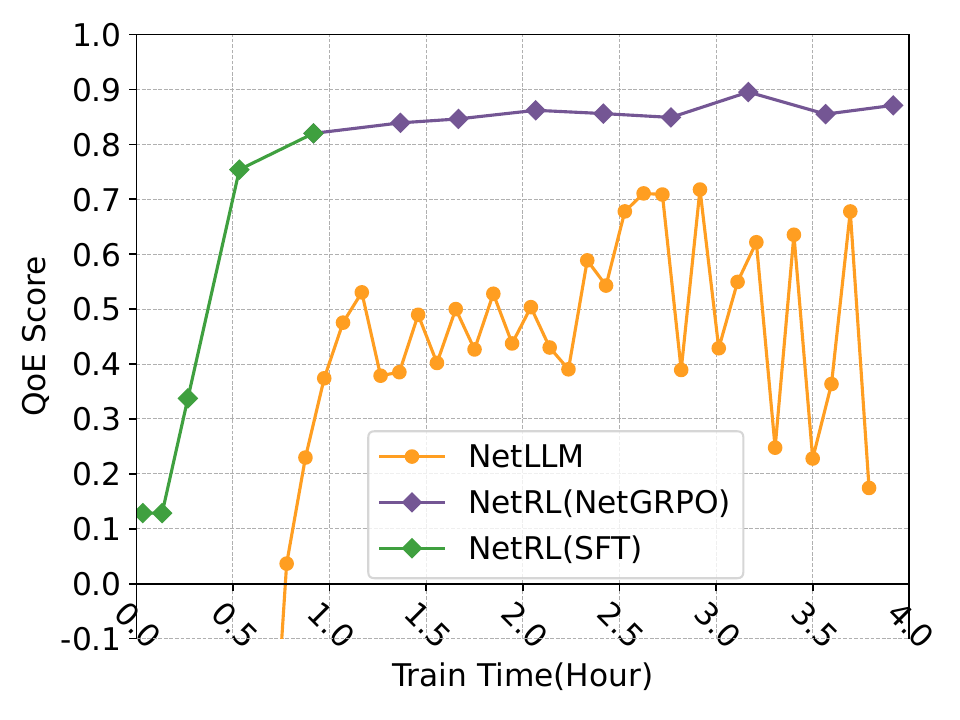}
    \caption{ABR training dynamics for NetLLM, NetRL SFT, and NetGRPO.}
    \label{fig:stability}
\end{figure}

Fig.~\ref{fig:stability} shows that SFT reaches approximately 0.82 QoE within one hour, after which NetGRPO steadily improves it to 0.89 over 2--3 hours. Multi-path evaluation reduces action-value noise, while group-relative normalization, clipping, and KL regularization constrain policy changes. Although tree generation lengthens each iteration, these more reliable updates reduce regressions and allow convergence within a relatively short time. NetLLM's offline checkpoints instead fluctuate substantially and plateau near 0.72 because its imitation objective lacks an explicit mechanism for constraining changes in the large policy.

\subsection{Foundation Model Selection and Computational Cost}

\begin{figure}[t]
    \centering
    \includegraphics[width=0.64\linewidth]{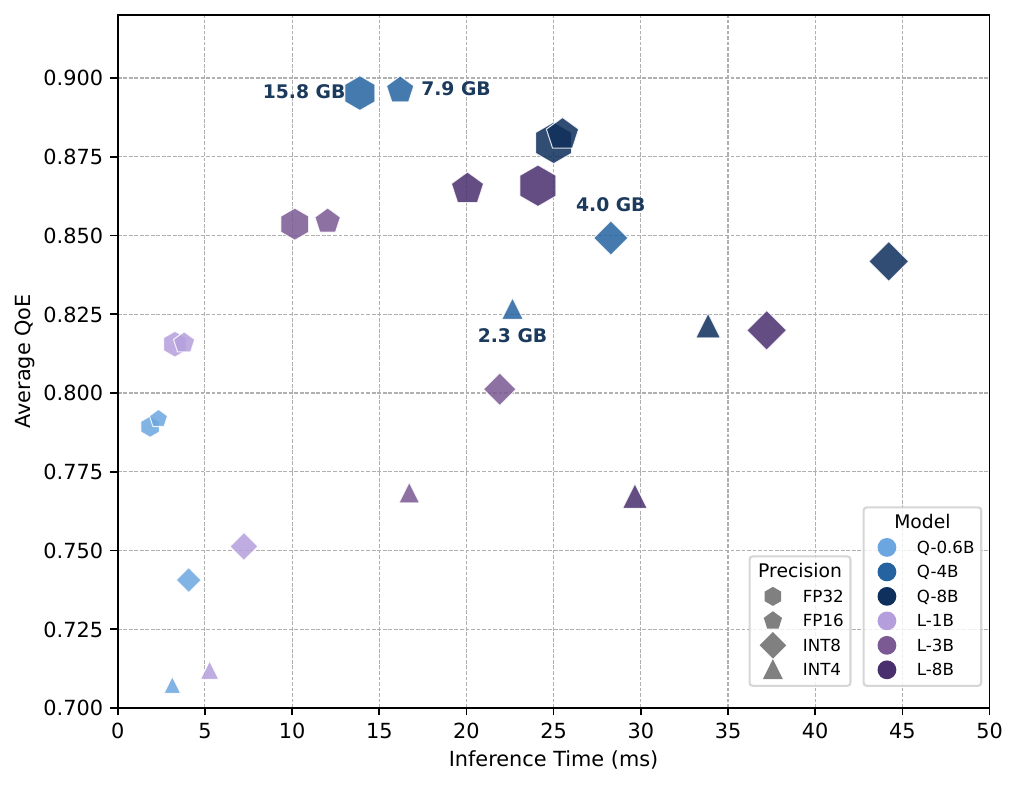}
    \caption{QoE and inference latency of candidate foundation models on ABR. Marker size indicates GPU memory usage.}
    \label{fig:model_choose}
\end{figure}

\textbf{Foundation Model Selection.} On an NVIDIA A100 GPU, we compare Qwen3-0.6B/4B/8B and Llama3.2-1B/3B/8B. Fig.~\ref{fig:model_choose} shows that Qwen3-4B provides the best ABR tradeoff, reaching 0.89 QoE at 15\,ms; Qwen3-8B increases latency to 25\,ms without improving QoE. We therefore select FP32 Qwen3-4B.

\textbf{Computational Requirements.} NetRL's 4.11B parameters require 15.6\,GB in FP32. FP16 uses 7.9\,GB without QoE degradation; INT8 and INT4 use 4.4\,GB and 2.8\,GB at a 5--8\% QoE cost. Tree generation increases an iteration from 16\,ms to 870\,ms but is training-only. SFT and NetGRPO take 3--4 hours on four A100 GPUs, while deployment requires one 15\,ms forward pass.

\textbf{Confidence-Aware Tree Budgeting.} On 80 held-out ABR states, the controller stops 40\% of searches at depth 3 and otherwise extends 2.38 of 6 root actions on average. It matches every fixed-depth action with zero value regret while reducing transitions by 53.3\% and search time by 44.1\%. Across three paired 100-update runs, it reduces transitions by $54.6{\pm}2.8\%$, search time by $45.1{\pm}2.4\%$, and end-to-end time by $27.1{\pm}2.5\%$, while preserving validation QoE (0.993$\pm$0.005 vs.\ 0.991$\pm$0.004) and the bad quality ratio (3.17\% for both).

\subsection{Generalization}

We freeze the ABR model trained on FCC traces and evaluate it without retraining on two independent synthetic trace sets drawn from more challenging distributions.

\begin{figure}[t]
    \centering
    \subfloat[Unseen Setting 1]{\label{fig:unseen_setting_1}\includegraphics[width=0.48\columnwidth]{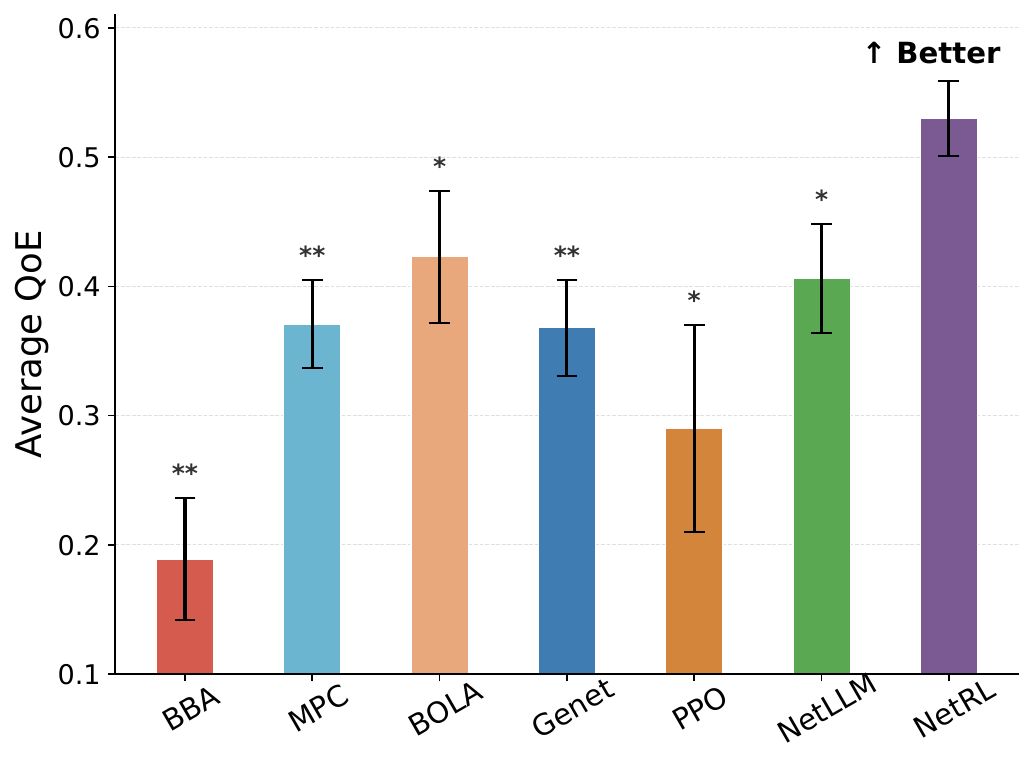}}%
    \hfill
    \subfloat[Unseen Setting 2]{\label{fig:unseen_setting_2}\includegraphics[width=0.48\columnwidth]{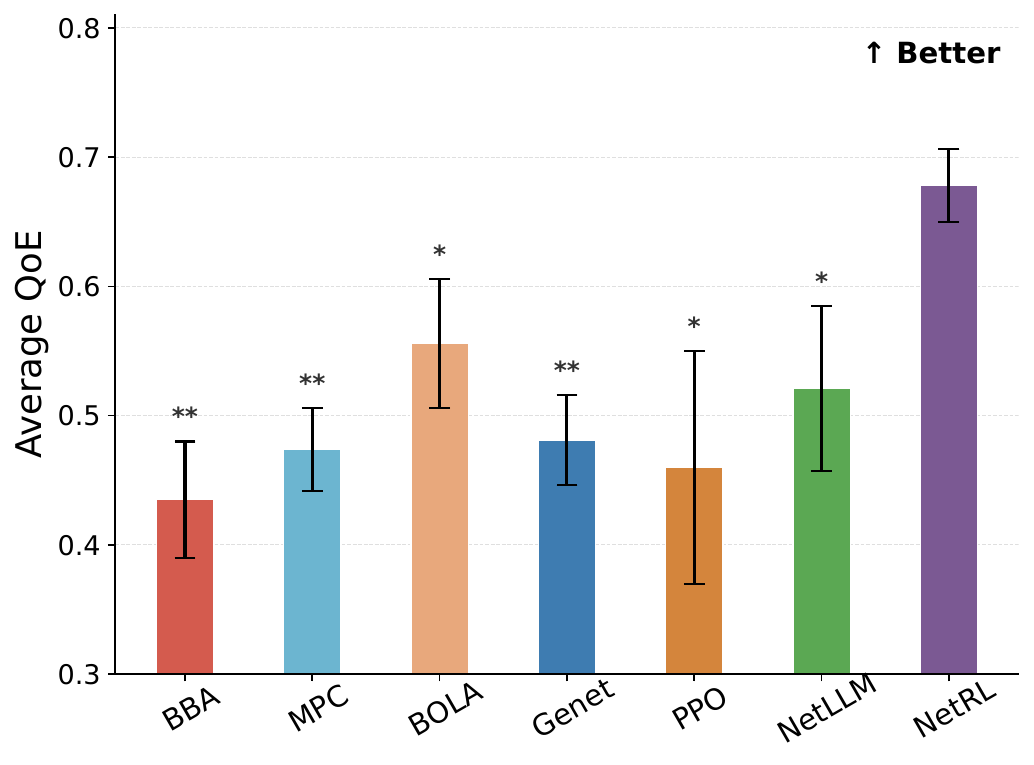}}%
    \caption{ABR performance on two independent out-of-distribution trace sets without retraining.}
    \label{fig:generalization}
\end{figure}

As shown in Fig.~\ref{fig:generalization}, NetRL achieves 0.53 and 0.68 QoE, outperforming NetLLM by 32.5\% and 30.7\%. All methods degrade under the more demanding conditions, but NetRL remains best in both settings. BOLA falls from 0.72 on the standard test set to 0.42 and 0.56 because its fixed rules adapt poorly as conditions deteriorate. Thus, NetRL generalizes to unseen, more challenging trace distributions and can better accommodate diverse production conditions.

\section{Discussion}
\label{sec:discussion}
\textbf{Production Deployment.} Deployment requires only one 15\,ms policy forward pass. FP16 uses 7.9\,GB without degrading QoE, while INT4 uses 2.8\,GB with a 5--8\% QoE reduction. Distillation, acceleration, and inference offloading provide further paths to resource-constrained deployment.

\textbf{Limitations and Future Work.} NetRL currently targets discrete actions and simulation. Future work will study continuous spaces, physical testbeds, and production workloads.

\section{Conclusion}
\label{sec:conclusion}
We presented \textbf{NetRL}, combining a universal mixed-embedding interface with NetGRPO's confidence-aware multi-path evaluation and stable online optimization. Across adaptive bitrate streaming, load balancing, and congestion control, NetRL achieves the best overall performance among the evaluated methods, robust tail behavior, stable convergence, and generalization to unseen traces. Confidence-aware budgeting further reduces ABR training cost without degrading validation performance.

\section*{Use of AI Disclosure}
The authors used DeepSeek and OpenAI Codex for language editing and code assistance. All content and results were independently verified by the authors.

\clearpage
\bibliographystyle{IEEEtran}
\bibliography{acmart}

\end{document}